
%
\input harvmac
\def\figflag{I}
\def\psheader#1{}


\font\blackboard=msbm10 \font\blackboards=msbm7
\font\blackboardss=msbm5
\newfam\black
\textfont\black=\blackboard
\scriptfont\black=\blackboards
\scriptscriptfont\black=\blackboardss
\def\blackb#1{{\fam\black\relax#1}}



%
 
\def\BR{{\blackb R}} \def\BI{{\blackb I}}
\def\BZ{{\blackb Z}}

\def\tfig#1{Fig.~\the\figno\xdef#1{Fig.~\the\figno}\global\advance\figno by1}
\def\figI{I}
%
\newdimen\tempszb \newdimen\tempszc \newdimen\tempszd \newdimen\tempsze
\ifx\figflag\figI
\input epsf
%
\def\epsfsize#1#2{\expandafter\epsfxsize{
 \tempszb=#1 \tempszd=#2 \tempsze=\epsfxsize
     \tempszc=\tempszb \divide\tempszc\tempszd
     \tempsze=\epsfysize \multiply\tempsze\tempszc
     \multiply\tempszc\tempszd \advance\tempszb-\tempszc
     \tempszc=\epsfysize
     \loop \advance\tempszb\tempszb \divide\tempszc 2
     \ifnum\tempszc>0
        \ifnum\tempszb<\tempszd\else
           \advance\tempszb-\tempszd \advance\tempsze\tempszc \fi
     \repeat
\ifnum\tempsze>\hsize\global\epsfxsize=\hsize\global\epsfysize=0pt\else\fi}}
\epsfverbosetrue
\psheader{fig.pro}       
\fi
%

%
%
%
%

\def\ifigure#1#2#3#4{
\midinsert
\vbox to #4truein{\ifx\figflag\figI
\vfil\centerline{\epsfysize=#4truein\epsfbox{#3}}\fi}
\narrower\narrower\noindent{\bf #1:} #2
\endinsert
}


%

\def\del{\partial}
\def\fourpt{\hbox{{$\rangle \kern-.25em \langle$}}} 
\def\tree{\hbox{{$\rangle \kern-.5em - \kern-.5em \langle$}}}
 %

\def\H#1#2{{\rm H}^{#1}(#2)} 
\def\CZ{{\cal Z}}
\def\ex#1{{\rm e}^{#1}}                 
\def\dd{\mskip 1.3mu{\rm d}\mskip .7mu} 

\def\RPtwo{\BR {\rm P}^2}
\noblackbox

\Title{\vbox{\hbox{PUPT--1324}\hbox{\tt hepth@xxx/9205100}}}
{A note on the 3D Ising model as a string theory$^\star$}

\centerline{Jacques Distler}\smallskip
\centerline{Joseph Henry Laboratories}
\centerline{Princeton University}
\centerline{Princeton, NJ \ 08544 \ USA}
\bigskip\bigskip

\footnote{}{{\parindent=-10pt\par $\star$
\vtop{
\hbox{Email: {\tt distler@puhep1.princeton.edu} .}
\hbox{Research supported by NSF grant PHY90-21984.}
     }     }}

It has long been argued that the continuum limit of the 3D Ising model is
equivalent to a string theory. Unfortunately, in the usual starting point for
this equivalence -- a certain lattice theory of surfaces -- it is not at all
obvious how to take the continuum limit. In this note, I reformulate the
lattice
theory of surfaces in a fashion such that the continuum limit is
straightforward.
I go on to discuss how this new formulation may overcome some fundamental
objections to the notion that the Ising model is equivalent to a string theory.
In an appendix, I also discuss some aspects of fermion doubling, and the
lattice
fermion formulation of the 2D Ising model.

\Date{5/92}                 

\lref\onsager{L. Onsager, Phys. Rev. {\bf 65} (1944) 177.}
\lref\kacward{M. Kac and J. C. Ward, Phys. Rev. 88 (1952) 1332.}
\lref\feynman{R. Feynman, {\it Statistical Mechanics, A Set of Lectures}, W. A.
Benjamin (1972).}
 \lref\sherman{S. Sherman, J. Math. Phys. {\bf 1} (1963) 202,
{\it ibid} {\bf 4} (1963) 1213.}
\lref\potts{R. Potts and J. C. Ward, Prog. Theor. Phys. {\bf 13}
(1955) 38.}
\lref\hurst{C. Hurst and H. Green, J. Chem. Phys. {\bf 33} (1960) 1059.}
\lref\schultz{T. Schultz, D. Mattis and E. Lieb, Rev. Mod. Phys. {\bf 36}
(1964) 856.}
\lref\AGMV{L. Alvarez-Gaum\'e, G. Moore and C. Vafa, Comm. Math
Phys. {\bf 106} (1986) 40.}
\lref\FSS{E. Fradkin, M. Srednicki and L. Susskind, Phys. Rev. {\bf D21} (1980)
 2885.}
\lref\PolyO{A. Polyakov, Phys. Lett. {\bf 103B} (1981) 211.}
\lref\Polybook{A. Polyakov, {\it Gauge Fields and Strings}, Harwood Academic
Publishers (1987).}
\lref\DotPol{Vl. Dotsenko and A. Polyakov, {\it in} Advanced
Studies in Pure Math 15\semi
Vl. Dotsenko, Nucl. Phys. {\bf B285} (1987) 45.}
\lref\CFWI{A. Casher, D. F\oe rster and P. Windey, Nucl. Phys. {\bf B251}
(1985) 29\semi
C. Itzykson, Nucl. Phys. {\bf B210} (1982) 477.}
\lref\sedrakyan{A. Sedrakyan,
Phys. Lett.. {\bf 137B} (1984) 397\semi
A. Kavalov and A. Sedrakyan, Phys. Lett. {\bf 173B} (1986) 449\semi
A. Kavalov and A. Sedrakyan, Nucl. Phys. {\bf B285} (1987) 264.}
\lref\DotDot{V. Dotsenko and Vl. Dotsenko, Adv. Phys. {\bf 32} (1983) 129.}
\lref\kutasov{D. Kutasov and N. Seiberg,   Nucl. Phys. {\bf
B358} (1991) 600.}
\lref\finite{H. Bl\"ote, J. Cardy , and M. Nightingale, Phys.
Rev. Lett.. {\bf 56} (1986) 742\semi
I. Affleck, Phys. Rev. Lett.. {\bf 56} (1986) 746.}
\lref\milnor{J. Milnor and J. Stasheff, {\it Characteristic Classes}, Princeton
University Press (1974)}
\lref\moise{E. Moise, {\it Geometric Topology in Dimensions 2 and 3},
Springer-Verlag (1977)}
\lref\spinstr{M. Atiyah, R. Bott and A. Shapiro, Topology {\bf 3} (1964) 3.}
\lref\whitney{M. Golubitsky and V. Guillemin, {\it Stable Mappings and Their
Singularities}, Springer-Verlag (1973).}
\lref\rohm{B. Grinstein and R. Rohm, Comm. Math. Phys. 106 (1986) 40.}
\lref\largeorder{P. Ginsparg and J. Zinn-Justin, Phys. Lett.. {\bf B255} (1991)
189\semi
S. Shenker, {\it in} 1990 Carg\`ese Workshop on Random Surfaces, Quantum
Gravity and Strings, RU90-47.}
\lref\sedratalk{A. Sedrakyan, talk at Princeton, Fall, 1991.}
\lref\kurzepa{For a discussion, see T. Jarosewicz and P. Kurzepa, Ann. Phys.
{\bf 210} (1991) 255.}
\lref\GS{M. Green and J. Schwarz, Phys. Lett. {\bf 151B} (1985) 21.}
\lref\CBDM{
Y. Choquet-Bruhat and C. DeWitt-Morette, ``{\it Analysis, Manifolds and
Physics}, part II: Applications", North Holland (1989).}
\newsec{Introduction}

The 2D Ising model was first solved by Onsager \onsager\ in the 1940's, and has
been reformulated and re-solved by many others subsequently (some references
relevant to my discussion are
\refs{\kacward,\feynman,\sherman,\potts,\hurst,\schultz}). It has been a rich
and
fruitful source of insights into 2D critical phenomena.

The 3D Ising model has, so far, resisted an exact solution. Though much is
known about its critical behaviour, both analytically and numerically, many
have hoped that it might yet yield, if not to an exact solution, then at least
to a closed-form calculation of its critical exponents.
One of the most elegant suggestions for reformulating the 3D Ising model has
been to try to recast it as a string theory, that is, as a theory of surfaces
immersed in 3 dimensions.

Despite much work on this subject
\refs{\FSS,\PolyO,\Polybook\DotPol,\CFWI,\sedrakyan}, not much progress has
been
made. In this paper, I will reconsider the traditional approach to  recasting
the 3D Ising model as a string theory and try to suggest where it runs into
difficulties. I will then suggest an alternate approach which has the promise
of
overcoming these difficulties.

Anyone who tries to to convince you that the continuum limit of the 3D Ising
model is equivalent to some string theory had better be prepared to address at
least two key objections:
\item{$\bullet$}String theory would seem to have far too many degrees of
freedom to be equivalent to  what is, after all, a {\it field theory} in 3
dimensions.
\item{$\bullet$}Intimately related \kutasov\ to this large density of states is
the fact that string theory generically has a tachyon in its spectrum, which is
surely {\it not} a property of the 3D Ising model. In practice, the tachyon
leads to a divergence of the string integrand near surfaces with long thin
tubes. In other words, these configurations tend to dominate the functional
integral, whereas we expect that near the critical point of the Ising model,
domains of all sizes become important.

These problems are {\it very generic} and are more or less independent of the
world-sheet theory (so long as it has a large enough number of degrees of
freedom). Since precisely the same problems are currently faced by noncritical
string theory (for $D>1$), it is very interesting to see how the 3D Ising
model manages to solve them. This is the main motivation behind the present
work.

The plan of this paper is as follows. In section 2, I review the 2D Ising model
from a point of view which will generalize to 3 dimensions. In section 3, I
discuss the 3D Ising model. I discuss the  approach adopted in previous
attempts to recast it as a string theory and suggest an alternate approach,
where taking the continuum limit is more straightforward. In section 4, I
return
to the problems just mentioned, and discuss how, in this new approach to the 3D
Ising model, they might be solved.

There are three appendices. Appendix A is devoted to the lattice fermion
representation of the 2D Ising model, and how it avoids fermion doubling.
Appendix B contains a brief speculation on the subject of the equivalence
between the Nambu-Goto and Polyakov strings in 3 dimensions. Appendix C
is a review of the definition of fermions on nonorientable manifolds.

\newsec{Review of the 2-D Ising model}

Let us start our discussion by recalling the situation in the 2-D Ising model.
The partition function is obtained by summing over all configurations of spins
$\sigma_i=\pm1$ on the sites of the lattice
\eqn\eZtwo{\CZ[\beta]=\sum_{\{\sigma\}} \ex{-\beta\sum_{<ij>}
(1-\sigma_i\cdot\sigma_j)}\quad. }
The action $S=\sum_{<ij>}
(1-\sigma_i\cdot\sigma_j)$ only receives contributions from links across which
the neighbouring spins are anti-aligned. If we have a domain of up spins
($\sigma=+1$) next to a domain of down spins ($\sigma=-1$), the contribution
to the action is proportional to the length of the boundary separating the two
domains. Thus we can contemplate replacing \eZtwo\ by a sum over {\it curves}
(on the dual lattice) weighted by their lengths \feynman
\eqn\eZtwoa{\CZ[\beta]=2Z[\beta]{\buildrel ?\over=}
2\sum_{\{\gamma\}}\ex{-2\beta L[\gamma]}}
The factor of 2 in this formula simply accounts for the $\BZ_2$ symmetry of the
 Ising model, in which flipping all of the spins $\sigma_i\to-\sigma_i$
produces the same configuration of curves.
To make this formula correct, we should first of all restrict ourselves to
closed (but not necessarily connected) curves $\gamma$ which traverse
each link of the (dual)
 lattice only {\it once} (or zero times). This restriction
to so-called {\it unicursal} curves will be relaxed later, but for the moment
it is still not enough to make the equality \eZtwoa\ correct.

The problem is that the curve $\gamma$ can self-intersect.
If we consider the configuration of 4 spins shown in \tfig\fone{a}, we see that
the corresponding curve $\gamma$ has four links which apparently meet at a
point. There are three ways to resolve
this singular point (\fone b,c,d). That is to say, there are three different
immersed curves whose images involve the same four links. Counting immersed
curves as in \eZtwoa, we would count each of these configurations separately.
This is
{\it overcounting} since they all correspond to the same configuration of
spins.
We can fix this by introducing the intersection number $n(\gamma)$ which counts
the number of intersections of the immersed curve $\gamma$. \fone d
contributes 1 to $n(\gamma)$, while \fone b,c do not intersect, and so
contribute 0 to $n(\gamma)$. Instead of \eZtwoa, consider
\eqn\eZtwob{\CZ[\beta]=2Z[\beta]= 2\sum_{\{\gamma\}}(-1)^{n(\gamma)}\ex{-2\beta
L[\gamma]}\quad.} This counts \fone d with a minus sign relative to \fone b,c.
Two of the three configurations cancel leaving a {\it net} of one
configuration, which agrees with the sum over spins.

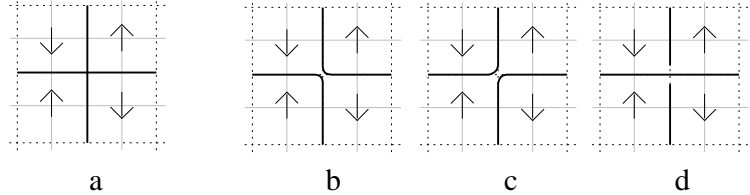
\ifigure\fone{A configuration of spins, and three different
immersed curves corresponding to it}{fone.eps}{1.07}

Having introduced the factor of $(-1)^{n(\gamma)}$, we can relax the
requirement that the curve $\gamma$ be unicursal. This is good because the
unicursal restriction is an extremely nonlocal one, and would like to replace
it with something local. Let us define an {\it admissible} curve to be a closed
(not necessarily connected) curve on the lattice which does not retrace itself.
That is, no link is traversed twice in {\it successive steps}. The three
configurations \tfig\ftwo a,b,c are allowed; the configuration \ftwo d is not
allowed.

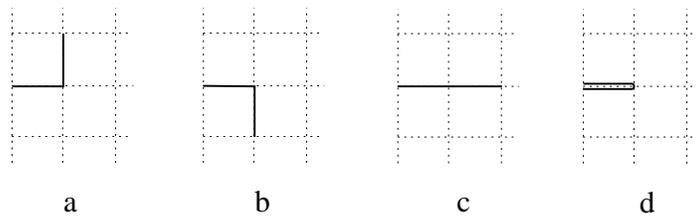
\ifigure\ftwo{Three admissible configurations and one inadmissible
one}{ftwo.eps}{1.14}

Consider the admissible configuration in \tfig\fthree a. The central link is
occupied twice. This configuration corresponds to four different admissible
curves (\fthree b,c,d,e). But with the sign factor in \eZtwob, two of these
configurations get counted with a plus sign and two with a minus. Thus they
cancel,
leaving only (\tfig\ffour) as the curve $\gamma$ corresponding to this
configuration of spins.

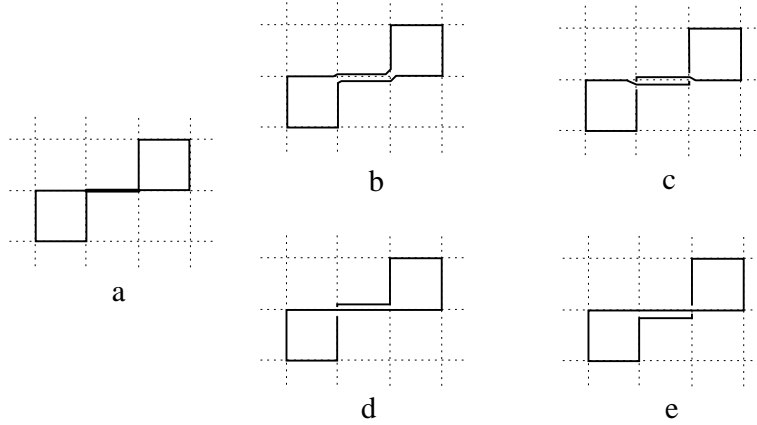
\ifigure\fthree{An admissible configuration and its
resolutions.}{fthree.eps}{2.27}

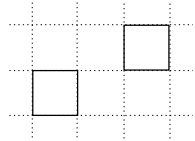
\ifigure\ffour{The admissible curve $\gamma$ which contributes to the partition
function.}{ffour.eps}{.757}

In fact, we claim
\eqn\eident{\sum_{\left\{{unicursal\atop\gamma}\right\}}
(-1)^{n(\gamma)}\ex{-2\beta L[\gamma]}=
\sum_{\left\{{admissible\atop\gamma}\right\}}
(-1)^{n(\gamma)}\ex{-2\beta L[\gamma]}}
To prove this, we need to show that the cancellation seen in the above example
is a general phenomenon whenever any link is multiply-occupied.
Consider a vertex at which four links, occupied $k_1$ through $k_4$ times
respectively, come together (\tfig\ffive a). Let $k_1$ be the largest of the
$k_i$.

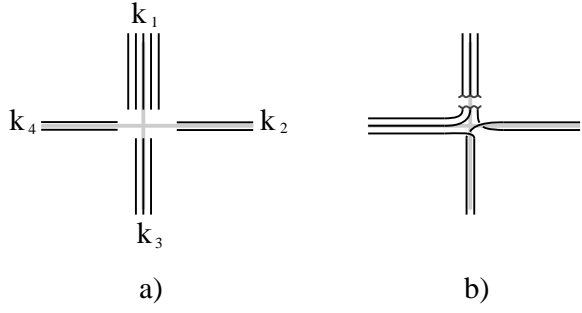
\ifigure\ffive{A vertex at which four multiply-occupied links
meet}{ffive.eps}{1.6}

The easy case to handle is when
$k_1=k_2+k_3+k_4$. In this case, each line from links $2,3,4$ {\it must} be
connected to  a line from link 1. The number of different ways of doing this
is given by the number of permutations of the $k_1$ lines on link 1.
The sign factor associated to a particular way of connecting the lines is
given by the {\it signature} of the corresponding permutation,
$$(-1)^n=\sigma(g),\qquad g\in S_{k_1}\quad.$$
Summing over all possible ways of connecting the lines amounts to summing over
$S_{k_1}$ the group  of permutations of $k_1$ objects. But it is elementary
that
 $$\sum_{g\in S_{k_1}}\sigma(g)=\cases{0&$k_1>1$\cr 1&$k_1=1$\ .\cr}$$
So we get zero unless $k_1=1$, in which case, one of $k_{2,3,4}=1$, and the
other two are zero.

The harder case is when $k_1+2p=k_2+k_3+k_4$. In this case, we connect up
$2p$ lines among $k_{2,3,4}$, leaving $k_2+k_3+k_4-2p$ lines to connect to
$k_1$. (Later, we will sum over the different ways of making the first $p$
connections.) But this reduces the computation of the sign factor to an
overall sign (determined by our choice of the first $p$ connections) multiplied
by the same $\sigma(g)$ (see \ffive b). Again, summing over all $g\in S_{k_1}$
gives zero unless $k_1=1$.

This argument, a much-simplified version of the one appearing in \sherman,
allows us to replace unicursal curves $\gamma$ by admissible curves in \eZtwob,
\eqn\eZtwoc{Z[\beta]=
\sum_{\left\{{admissible\atop \gamma}\right\}}(-1)^{n(\gamma)}\ex{-2\beta
L[\gamma]}={\rm exp}\biggl[\sum_{\left\{{connected\atop{admissible\atop
\gamma}}\right\}}(-1)^{n(\gamma)}\ex{-2\beta L[\gamma]}\biggr]\quad.}
Note that the sum over admissible curves has another advantage -- it
exponentiates -- as a sum over unicursal curves would not.

{}From here it is relatively easy to see that the continuum limit of this
theory
is a theory of free  majorana fermions. The most direct way is to define a
theory of lattice fermions \refs{\hurst,\schultz}\ whose partition function is
reproduced by the
same diagrammatic expansion as \eZtwob. Some issues related to this approach
are discussed in Appendix A.
More intuitively, we
recognize \eZtwoc\ as the (Euclidean) partition function for a 1+1 dimensional
quantum field theory, in a first-quantized formalism (familiar to any string
theorist). The particles do not interact, but they do have statistics. Every
time two particle trajectories cross (the only way particles can be exchanged
in 1+1 dimensions) we get  a minus sign -- so the particles are fermions.

The connection with free fermions becomes even more clear when we consider the
theory on a lattice with periodic boundary conditions \potts. On a finite
lattice  with periodic boundary conditions, the sum over curves still
overcounts configurations.  Consider a loop which winds around one of the
homology cycles of the torus. This doesn't correspond to any configuration of
spins. More generally, $\gamma$ can have total winding number $p_t$ in the
horizontal direction and $q_t$ in the vertical direction. Only $(p_t,q_t)$
{\it even} correspond to physically realizable spin configurations.

The solution is simple. First introduce ``independent" horizontal and vertical
couplings $\beta_h$, $\beta_v$. Define
\eqn\eZbetahv{Z(\beta_h,\beta_v)=\sum_{\{\gamma\}}(-1)^{n(\gamma)}
\ex{-2(\beta_hL_h[\gamma]+\beta_vL_v[\gamma])}}
where $L_{h,v}[\gamma]$ is the number of horizontal (vertical) links in
$\gamma$ ($L[\gamma]=L_h+L_v$). Our old partition function is just
$Z[\beta]=Z(\beta,\beta)$. But now consider
$$Z(\beta+i\pi/2,\beta)=\sum_{\{\gamma\}}(-1)^{n(\gamma)}\ex{-2\beta L[\gamma]}
(-1)^{p_t(\gamma)N_h}$$
where $N_{h,v}=$ horizontal (vertical) length of the lattice ($N=N_hN_v$).
Similarly,
$$Z(\beta,\beta+i\pi/2)=\sum_{\{\gamma\}}(-1)^{n(\gamma)}\ex{-2\beta L[\gamma]}
(-1)^{q_t(\gamma)N_v}\quad.$$
To reproduce the Ising partition function on a periodic lattice, we simply
need to sum
\eqna\eZtwod
$$\eqalignno{\CZ[\beta]&=\half[Z(\beta,\beta)+Z(\beta+i\pi/2,\beta)\cr
&\qquad+Z(\beta,\beta+i\pi/2)+Z(\beta+i\pi/2,\beta+i\pi/2)]&\eZtwod a\cr
&=2 \sum_{\{\gamma\}}\half\Bigl(1+(-1)^{p_t(\gamma)N_h}\Bigr)
\half\Bigl(1+(-1)^{p_t(\gamma)N_h}\Bigr)(-1)^{n(\gamma)}\ex{-2\beta
L[\gamma]}&\eZtwod b\cr}$$
Following \potts, we assume the $N_{h,v}$ are both {\it odd}, and we see that
$\eZtwod{b}$ projects onto $q_t,p_t$ both even. But the terms in the sum
$\eZtwod{a}$ are clearly recognizable\footnote{${}^\star$}{For a dirac fermion,
one can follow \AGMV\ and represent the different spin structures by coupling
the
fermion to a flat $U(1)$ bundle. For a majorana fermion on the lattice, the
same trick is available, provided we take the holonomy across each link of the
lattice to be $\pm1$.}\ as the sum over spin structures of the fermion. (This
connection is made explicit in Appendix A). At the
critical point, the contribution of the odd spin structure,
$Z(\beta,\beta)$, vanishes.

\newsec{On to 3-D $\dots$}

The lesson that can be abstracted from the previous section can be summarized
as follows. We wanted to recast the 2-D Ising model as a theory of
immersed curves. However, if we are not careful, the sum over immersed curves
overcounts configurations. The point is that there are two natural topologies
that one can associate with a curve immersed in $\BR^2$. One is the intrinsic
topology of the curve. The other is the topology inherited from $\BR^2$, the
{\it extrinsic} topology. These are in general different.
Points which are far apart in the intrinsic topology
may be close together in the extrinsic topology. What is more, there are many
possible intrinsic topologies which give rise to the same extrinsic topology.
But the Ising model action depends only on the
the extrinsic geometry of the immersed curve, and is insensitive to
the intrinsic topology. Thus we needed to introduce a topological term
$(-1)^{n(\gamma)}$ which distinguished between the different intrinsic
topologies
 corresponding to a given extrinsic geometry and introduced cancellations
between them.

In three dimensions, the boundaries between domains are closed surfaces, not
curves. Again, we restrict ourselves at first to unicursal surfaces --
those that occupy each plaquette (of the dual lattice) at most once.
We wish to rewrite the Ising partition function as a sum over surfaces
weighted by their area
\eqn\eZthreea{\CZ[\beta]=2Z[\beta]{\buildrel ?\over=}
2\sum_{\{\Sigma\}}\ex{-2\beta A[\Sigma]}}
but, again, this overcounts configurations because the surface may have many
intrinsic topologies corresponding to a given extrinsic topology.

One possible solution, suggested by Fradkin, Srednicki and Susskind \FSS\ and
elaborated by Polyakov and Dotsenko \DotPol\ and others \sedrakyan, is a simple
generalization of \eZtwob\ to 3 dimensions,
\eqn\eZthreea{\CZ[\beta]=2Z[\beta]=
2\sum_{\{\Sigma\}}(-1)^{L_{int}[\Sigma]}\ex{-2\beta A[\Sigma]}}
where $L_{int}$ is the length of the line of self-intersection of the surface
$\Sigma$ in lattice units.
The basic idea can be seen in \tfig\fsix. Here we have three configurations
(drawn as continuous surfaces), in which four sheets of the surface come
together along a line. One possible intrinsic topology corresponding to
this extrinsic topology is depicted in \fsix a. The surface is taken to
self-intersect along the whole line. The top sheet on one side
is connected to the bottom sheet on the other side, and vice versa. The
other two surface differ from the first only in how they are connected across
one (the left-most) link. In \fsix b, the top sheet on one side is connected
to the top sheet on the other. In \fsix c, it is connected to the bottom
sheet on the same side. In each case, I've drawn a dashed curve to indicate
the path taken by a ``bug" which starts out on the top sheet above. Note that
from the point of view of the bug ({\it i.e.}~in the intrinsic topology)
nothing
special happens as it crosses the line of self-intersection.

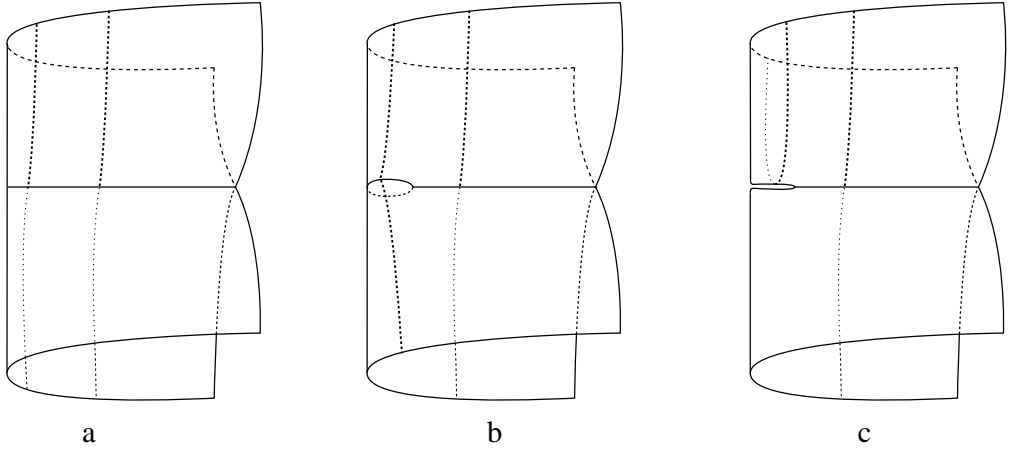
\ifigure\fsix{Three configurations, two of which must cancel in the sum over
surfaces}{fsix.eps}{2.38}

Since the surface \fsix a self-intersects along one more link of the lattice
than \fsix b,c, by the above rule it gets counted with a relative minus sign
in the sum over surfaces \eZthreea.

It was originally believed that the continuum limit of \eZthreea\ was the NSR
string in 3 dimensions \PolyO, and although Polyakov has subsequently suggested
other possibilities for what the continuum theory might be, nothing is
conclusively proven. Part of the difficulty is that  sign factor
$(-1)^{L_{int}[\Sigma]}$ oscillates very rapidly on the length scale of the
lattice spacing. It is therefore not obvious how one is supposed to take the
continuum limit. Also, the generalization from 2 to 3 dimensions of the
argument
that leads to fermions
is not as straightforward as one might think\kurzepa\ and
it is not at all clear that \eZthreea\ is equivalent to the NSR string.

One of the features that makes theories of surfaces in 3 dimensions so
much richer than that of curves in 2 dimensions is that the possibilities for
the intrinsic topologies are much more varied. In two dimensions, a curve
$\gamma$ was, in terms of its intrinsic topology, simply a disjoint union of
circles. The only feature which distinguished different intrinsic topologies
was the number of connected components. Surfaces have a much richer topological
classification and we will make use of this fact to replace the sign factor
$(-1)^{L_{int}[\Sigma]}$ with one that has a more obvious continuum limit.

Look again at \fsix. One thing is immediately evident is that one of the
surfaces that we have drawn (\fsix b) differs in its intrinsic topology from
the other two. Indeed, \fsix a,c are homotopic (they can be deformed
continuously into one another). To see exactly how \fsix b is different, we
need to make a short digression on the topology of 2-manifolds.

The topological classification of compact, connected 2-manifolds \moise\ is
completely given by the cohomology of the manifold with $\BZ_2$ coefficients.
The basic theorem is that one can choose a basis for $\H{1}{\Sigma,\BZ_2}$ such
that the generators have one of the following three intersection forms
\eqn\einter{\vbox{\halign{
#\quad&
\hfil $#$&$#$\hfil\quad&\quad #\quad&\hfil $#$&$#$\hfil\quad&\quad #\quad&
\hfil $#$&$#$\hfil\cr
i)&\alpha_i\cup\beta_j=&\delta_{ij}&ii)&\alpha_i\cup\beta_j=&\delta_{ij}&iii)&
\alpha_i\cup\beta_j=&\delta_{ij}\cr
&&&&\gamma\cup\gamma=&1&&\gamma\cup\gamma=&1\cr
&&&&&&&\gamma\cup\tilde\gamma=&1\cr
}}}
with all other cup products being zero.
The indices $i,j$ to run from 1 to $g$, where $g$ is the number of
``handles". The dimensions of $\H{1}{\Sigma,\BZ_2}$ in the three cases are,
respectively, $2g$, $2g+1$, and $2g+2$. For $g=0$, these correspond to
the sphere, the
projective plane, and the Klein bottle.

The Stiefel-Whitney classes of $\Sigma$ are easily computed from Wu's formula
\milnor, and the above cohomology ring:
\eqn\eSW{\vbox{\halign{
#\quad&
\hfil $#$&$#$\hfil\quad&\quad #\quad&\hfil $#$&$#$\hfil\quad&\quad #\quad&
\hfil $#$&$#$\hfil\cr
i)&w_1=&0&ii)&w_1=&\gamma&iii)&w_1=&\tilde\gamma\cr
&w_2=&0&&w_2=&1&&w_2=&0\cr
}}}
The first Stiefel-Whitney class, $w_1$ is the obstruction to orientability of
$\Sigma$. The second Stiefel-Whitney class $w_2(\Sigma)$ is the
obstruction \rohm\ to defining fermions (or, more technically, to defining a
$pin$-structure \spinstr) on $\Sigma$. For closed surfaces, it is the reduction
modulo 2 of the Euler characteristic
$$w_2(\Sigma)=\chi(\Sigma)\quad {\rm modulo}\ 2\quad.$$

It is easy to see that the surface \fsix b has Euler characteristic (or
equivalently $w_2$) which differs by 1 from that of \fsix a or c.  In
particular,
\fsix b is non-orientable. To see this, consider the path of a ``bug" in
\tfig\fseven. As the bug traverses the dotted path, it comes back to itself
with
a reversal of orientation. The dotted path is essentially dual to the new
$\BZ_2$ cohomology cycle $\gamma$ introduced above.

\ifigure\fseven{A cycle on a non-orientable surface}{fseven.eps}{2.214}

Let us try to make use of
this difference in intrinsic topology to introduce the sign factors which we
need. Instead of \eZthreea, let us take
\eqn\eZthreeb{\CZ[\beta]=2Z[\beta]=
2\sum_{\{\Sigma\}}(-1)^{w_2(\Sigma)}\ex{-2\beta A[\Sigma]}} Actually, when one
wants to discuss disorder operators, {\it i.e.} when one wants to include
surfaces $\Sigma$ with boundaries, there are further sign factors which must be
included:
\eqn\eZthreec{Z[\beta,C=\del\Sigma]=\sum_{\{\Sigma\}}
(-1)^{w_1(C)+w_2(\Sigma)}\ex{-2\beta A[\Sigma]}} but we will, for the purposes
of this paper, mostly restrict our attention to closed surfaces.

Note that non-orientable surfaces are {\it inevitable} if we want to play this
game of cancelling surfaces with the same extrinsic geometry against each
other. If we did not include \fsix b (which happens to be non-orientable), we
would
have had no way of cancelling \fsix a,c to leave a net of one configuration.
This
was just as true of the old weighting scheme \eZthreea\ as for the new one
\eZthreeb. The only difference is that now we are trying to make direct use of
the fact.

As an aside, let me point out an apparent paradox.
As we have argued, \eZthreea\ necessarily
involves surfaces, like the projective plane, $\RPtwo$, which have
nonvanishing $w_2$. This means that there is an obstruction to (the standard
definition of) fermions
on such surfaces.
We can understand this obstruction physically in the simple case of
$\RPtwo$. $\RPtwo$ can be built by gluing together a disk and  a m\oe bius band
along
their common boundary (an $S^1$). We can imagine doing the fermion path
integral
on $\RPtwo$ in stages: compute the path integral on the disk to obtain a state
on the boundary $S^1$. Similarly, do the path integral on the m\oe bius band to
obtain a state on its boundary. Finally, we obtain the path integral on
$\RPtwo$ by taking the inner product of the state associated to the disk and
the
state associated to the m\oe bius band. It is easy to see that the state
associated to the disk is in the NS sector (the nontrivial spin structure on
$S^1$), whereas the state associated to the m\oe bius band is in the R sector
(the trivial spin structure on $S^1$). Since these are orthogonal, the only
consistent definition of the path integral is to take it to vanish identically.
(If, however, we allow insertions, and consider amplitudes involving the
insertion of an {\it odd} number of spin operators, we {\it can} obtain a
nonzero result.)
But this is not the behaviour one expects to find in the NSR string, where
one expects $\RPtwo$ to have a dilaton tadpole \GS, and it is certainly not the
behaviour we want in the Ising model, where surfaces like $\RPtwo$ are supposed
to cancel the contributions of other surfaces.

The resolution to this paradox is that
on nonorientable manifolds, there are two {\it inequivalent} definitions of
fermions. The ``standard" definition indeed has as its obstruction $w_2$. The
other definition, in turn, has as its obstruction, $w_2+w_1\cup w_1$. As can be
seen from \einter,\eSW, the latter obstruction vanishes identically on any
2-manifold. Clearly, the thing to do is to define the fermions in the
nonorientable NSR string using the second definition. The existence of these
two inequivalent definitions of fermions, though a ``known" result \CBDM,
does not seem to be widely appreciated, even among open string theorists. I
have
therefore included a short summary in Appendix C. In any case,  it is unlikely,
 even using the second definition of fermions, that \eZthreea\ is
related to the NSR string.

But let us return to our discussion of \eZthreeb. Superficially, this seems to
introduce cancellations of the desired sort, but does it work in detail? Does
the lattice sum over surfaces precisely reproduce the Ising model sum over
spins? In answering this question, we must first face up to the fact that we
actually need to specify a {\it rule} for reconstructing intrinsic surfaces
from
a collection of plaquettes. This rule must have two crucial properties:
\item{$\bullet$}It must be {\it local}. That is to say that the choice of how
we
connect together plaquettes at one link of the lattice cannot depend on distant
information (say, on choices made at distant links.)
\item{$\bullet$}It must be {\it intrinsic}.
That is, if we have four sheets coming together at a link (\tfig\feight),
we can connect sheet
1 to sheet 2 and sheet 3 to 4, or we can connect 1 to 4 and 2 to 3, or we can
connect 1 to 3 and 2 to 4. An intrinsic rule is one which is invariant under
permutation of the labels of the sheets.

For later notational convenience,
we will denote the three ways of connecting the sheets together by using a
 solid, dashed, or gray line to colour the link in question (\feight).

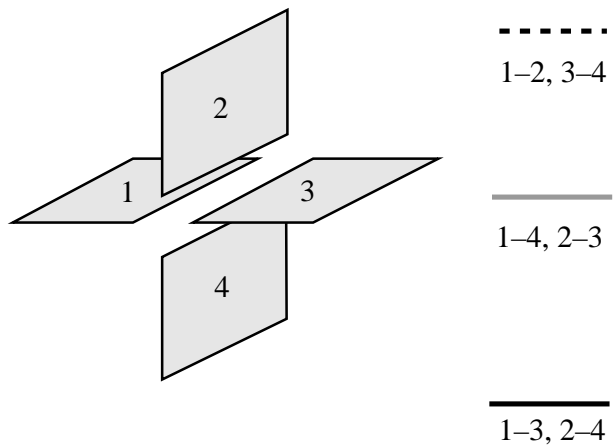
\ifigure\feight{Four sheets coming together on a line, and our notation for how
to connect them}{feight.eps}{2.365}

In the 2D case, there was only one possible intrinsic, local rule for turning a
collection of links into an immersed curve. We didn't even contemplate the
existence of possible alternatives\footnote{${}^\star$}{Note that if we were
willing to forego the requirements of locality and intrinsicness, there would
clearly exist rules which would eliminate the multiple counting -- without the
need to introduce the $(-1)^{n(\gamma)}$ factor.}. In the 3D case, we
must face up to the existence of a choice.

{}From the continuum point of view, the
lattice is providing a short-distance regulator for the theory. Different
``rules" correspond to slightly different lattice regulators for the theory
of surfaces. The notion of universality is simply that the details of the
lattice  regulator are irrelevant in the continuum limit. Because of the
tachyon divergence, as we shall discuss in the next section, such a naive
application of the principle of universality  needs to be checked carefully.

Before I state the rule which we shall use, let me motivate it heuristically.
 The lattice
acts as a regulator for the theory of surfaces basically by providing a minimal
length for cycles on $\Sigma$. We want the cutoff to be {\it shorter} for
cycles which carry the information about the orientability of $\Sigma$ ({\it
e.g.}~the cycle $\gamma$ in \fseven). As we have seen, these are cycles
$\gamma$ with
\eqn\egg{\gamma\cap\gamma=1\quad.}
 We will impose the condition that such cycles have length $\geq1$ in lattice
units, whereas all other cycles must have length $\geq2$. Note that, because of
the $\BZ_2$ grading, it is consistent to impose this shorter cutoff on cycles
 $\gamma$ satisfying \egg\ (the sum of two such cycles has length $\geq2$, as
it must). The choice of $1,2$ lattice units is somewhat arbitrary, and I could
double both of them, if you prefer, at the cost of somewhat complicating the
statement of the rule to be given presently.

The rule for what you are allowed to do at one link where four sheets of
$\Sigma$ come together will depend only on how the sheets are connected
immediately on either side of the link in question. That is to say, it will
depend on what is happening at ``nearest neighbour" links, but {\it not} on
more distant information. It is therefore a {\it local} rule.
The allowed configurations are depicted in \tfig\fnine. The circles represent
how the sheets are connected on either side of the link in question (using the
``colour" mnemonic introduced in \feight). Notice that, in some of the allowed
configurations, the way the sheets are connected changes at {\it half}-lattice
spacings. As I said above, this can be avoided at the cost of somewhat
complicating the statement of the rule. Basically, there are two
possible situations. Either the sheets are connected together in the same
fashion on either side of the link in question (the first three rows in
\fnine), or they are connected together in a different fashion (the last
three rows). By construction, the rule is invariant under permutation of the
colours. So all you need to see is one of the first three rows and one
of the last three, to reconstruct the whole table. There are, in each case,
five
possible choices for how to connect the sheets across the link in question. Of
the five, the first three lead to the same value of $w_2(\Sigma)$, and the last
two change $w_2(\Sigma)$ by 1\footnote{${}^\star$}{If you have trouble seeing
this, recall the definition of $\chi(\Sigma)$ as {\sl
vertices$-$edges$+$faces},
and count the {\it relative} contribution to $\chi(\Sigma)$ for each of the
entries of a given row of \fnine. $\Delta\chi$ for the last two entries in a
row
relative to the first three is {\it odd}.}. Thus, according to \eZthreeb, they
get counted with a minus sign relative to the first three. As in 2D, this
leaves
a {\it net} of one configuration which contributes.

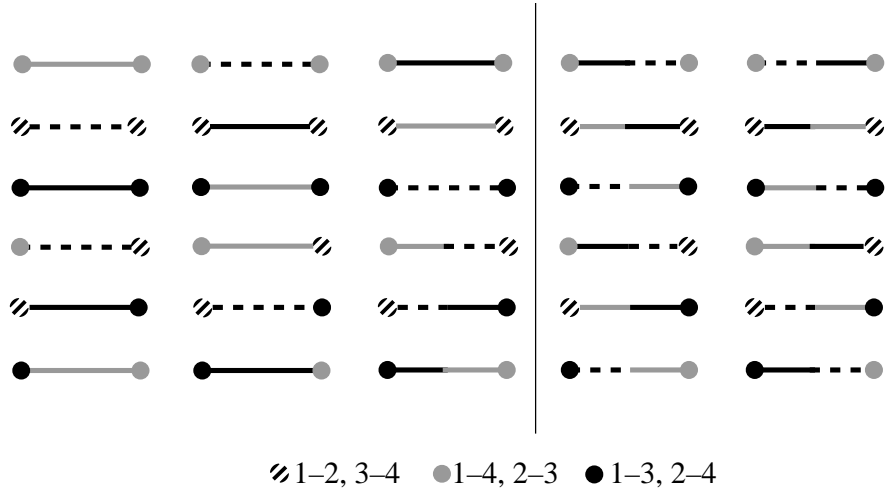
\ifigure\fnine{Allowed configurations on one link}{fnine.eps}{2.7}

This pretty much takes care of the proof that the sum \eZthreeb\ over unicursal
surfaces constructed according to the rule (\fnine) reproduces exactly the
Ising model partition function. The one point which should be clarified is the
notion of nearest neighbour link. We have four sheets coming together along a
curve. If the link under consideration is somewhere in the middle of that
curve, then it is clear what nearest neighbour means: the next link along the
curve. It may be, however, that the link in question is at the endpoint of the
curve (as in \fsix). But then, it is also unambiguous which sheets are
connected to which  ``to the left" of the link in question in \fsix. To
illustrate this, \tfig\fnineb\ contains two examples of configurations from
the  second row of \fnine, where the ``curve" in question is precisely one link
in length. On the right, you can see the path of a ``bug" on the corresponding
surfaces.

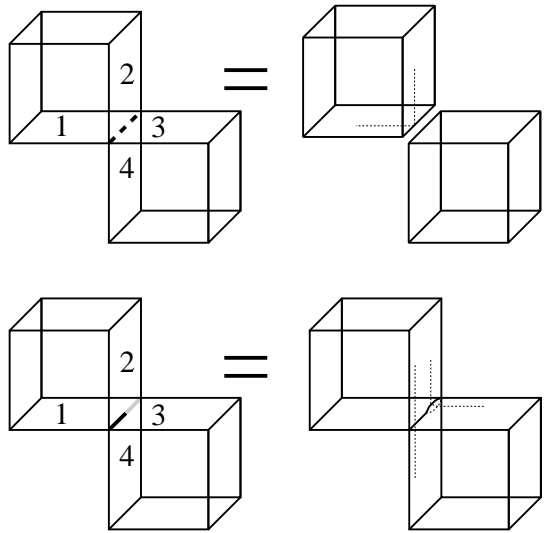
\ifigure\fnineb{Two configurations from the table above, and their
interpretation}{fnineb.eps}{2.82}

There is one case where the meaning of nearest neighbour {\it is} still
ambiguous,
namely when three sheets intersect each other at a point. A link adjoining this
vertex has five candidates for the role of ``nearest neighbour". The simplest
way to resolve this ambiguity is to define  the nearest neighbour as the link
straight ahead, since it is clearly singled out from the other four.

Having shown that the sum over unicursal surfaces reproduces the Ising
partition function, one would like to take the next step in this program and
relax the unicursal condition, by showing that the sum over admissible surfaces
with
weights given by \eZthreeb\ {\it equals} the sum over unicursal surfaces.
I have looked at simple examples in which this seems to work, but I have not
been able to construct a general proof along the lines of the one presented in
the last section for the 2D case.

Assuming that the replacement of unicursal by admissible surfaces goes through,
then the continuum limit of \eZthreeb\ is {\it obvious}. The sum over surfaces
on the lattice becomes a functional integral over immersed surfaces,
$A[\Sigma]$ simply becomes the Nambu-Goto action, and $w_2(\Sigma)$, being a
topological-invariant, is perfectly well-defined, both in the continuum and on
the lattice. Basically, we end up with the bosonic string in three dimensions
-- but with two crucial differences: 1) we sum over both oriented and
non-oriented 2-surfaces, and 2) we weight surfaces with odd Euler
characteristic with a {\it minus sign}.

\newsec{Comments on quantization}

When we recast the 2D Ising model as a sum over admissible curves, we
were fortunate to find that the free energy $W[\beta]$ was given by a sum
over {\it connected} curves {\it i.e.}~over maps from the circle into the
plane. Perhaps the most unattractive feature of recasting the 3D Ising model as
a string theory is that even the sum over connected surfaces involves a sum
over all genera of surface. This sum over genus is, in ordinary string
perturbation theory, controlled by a (purportedly) small coupling constant
$g_{st}$. In the Ising model, $g_{st}$ has magnitude $1$, since we want to
count domain boundaries of all genus with equal weight. This is an unpleasant
prospect to contemplate, since string perturbation theory is very likely
a divergent series\footnote{${}^\star$}{Of course, in the low-temperature
expansion, which has motivated our discussion, low-genus surfaces dominate
because they tend to be the surfaces of minimal area. However, we are precisely
interested in the theory near the critical point, where this area-suppression
breaks down.}.

We have somewhat improved this situation in the current formulation, since
effectively we have taken $g_{st}=-1$. If the large-order behaviour of the
perturbation series found in the matrix model \largeorder\ is any guide, this
is
enough to make the series Borel-summable\footnote{${}^\clubsuit$ }{For oriented
surfaces, the series goes like $(2g)! g_{st}^{(2g-2)}$. For non-oriented
surfaces, one has also odd powers of $g_{st}$, and the series goes like $n!
g_{st}^{(n-2)}$, which is Borel-summable for negative $g_{st}$.}.

But the summability of the perturbation series is a rather distant concern
compared with the more immediate problem of the tachyon divergence of {\it each
term} in the series. Here, too, introducing a negative string coupling
introduces
a measure of hope. Imagine introducing, in some uniform way, a cutoff on the
modular integration for each genus. This renders each term in the sum finite.
The contribution from each genus diverges as we send the cutoff to zero, but
now,
because there are explicit minus signs in the alternating sum, these
divergences
have the possibility of cancelling {\it between} different genera. Effectively,
when we  pinch a cycle as in \tfig\ften, we could be pinching an orientable or
a
non-orientable surface. These contribute with the same weight, but opposite
sign, to the sum over surfaces. If our cutoff is sufficiently clever, {\it all}
of the divergent pieces may cancel, leaving a result which is {\it finite} as
we
take the cutoff to zero. This may seem like outrageous handwaving, but the
lattice theory is an explicit realization of just such a cutoff. Hopefully, one
can find a more convenient one for the purposes of continuum calculations.

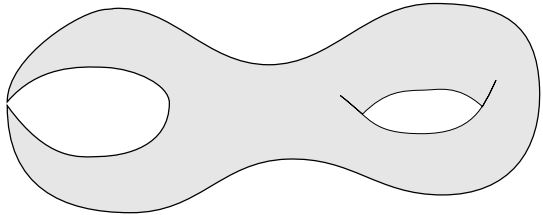
\ifigure\ften{A pinched surface could be the limit of an orientable or a
non-orientable surface, depending on the nature of the cycle(s)
pinched.}{ften.eps}{1.17}

Not only do the divergent terms cancel, which is to say that the contribution
of the tachyon in the pinched channel cancels out, but it is reasonable that
the contribution of many other states of the string cancel as well. In this
way, the strongly-coupled string theory may have many fewer propagating  states
than we expected. Indeed, it is only through such a drastic set of cancellation
that a string theory could ever reproduce the critical behaviour of the Ising
model, which is, after all, that of a field theory.

This mechanism, for cancelling divergences and reducing the number of
propagating states of a string theory is certainly novel. If, indeed, it works
(as it must, if the Ising model is to be described as a string theory), then
it is a telling example of just how different  strongly-coupled string dynamics
can be from that seen in perturbation theory. The details, of course, are
peculiar to D=3, and $|g_{st}|=1$. After all, the form of the string integrand
is D-dependent, and if this cancellation is to be realized for D=3, it will, in
general, not take place for other values of D. Similarly, the string coupling
constant is necessarily fixed by demanding that this cancellation take place.
Hence we expect that the dilaton, too, is projected out of the
spectrum\footnote{${}^\dagger$}{The dilaton had better be projected out for
another reason: were it not projected out, its presence would still lead to a
divergence in \ften.}.

With this vast truncation of the spectrum, string theory is clearly  not a very
{\it economical} description of the long-distance physics of the Ising model.
But then again, it is not a very economical description of elementary particle
physics either (which is also well-described as a field theory). The key
question we as string theorists are grappling with is how  string theory
with its good short-distance behaviour can describe the long-distance physics
that we see. The example of the Ising model should caution us that the answer
may be stranger than we think.

\appendix{A}{Undoubled lattice fermions, and the 2D Ising model}

As mentioned in the main text, the continuum limit of the 2D Ising model
is a free majorana fermion. This is most easily seen by formulating it as a
theory of lattice fermions, which is the subject of this appendix. When one
does so, one immediately is confronted with the problem of
fermion doubling\footnote{${}^\star$}{I'd like to thank P. Ginsparg for sharing
his insights into this subject}. Fermion doubling is a very generic feature of
lattice fermion theories. Should we be surprised  that the theory of
lattice fermions had as its continuum limit a {\it single} majorana fermion
which is massless at the critical point? How does this theory manage to avoid a
doubling of the spectrum, {i.e.}~why does  the continuum limit  not
produce two (or four) species of free majorana fermions?

If it did produce four majorana fermions, then the central charge, as measured
by the finite size scaling \finite\ would be $c=2$, rather than $c=1/2$. Other
critical exponents (such as the divergence of the correlation
length as we approach the critical point) would be the same for four decoupled
Ising models as for one, but the discrepancy in $c$ rules out this possibility.

We will examine this question both for the Ising model and for a closely
related model, the Ashkin-Teller model.

Let us begin with the Ising model. Following \hurst, one can introduce a
theory of lattice fermions  whose partition sum reproduces exactly the
diagrammatic expansion \eZtwob\ of the 2D Ising model. Introduce fermionic
variable $\psi_1,\psi_2,\psi_3,\psi_4$ living on the sites of the dual lattice
with the action \eqn\eSfermi{\eqalign{S=\sum_{\bf x}\
\ex{-2\beta}(\psi_1\hat\del_x\psi_3
+\psi_2\hat\del_y\psi_4)&+\psi_1\psi_2+\psi_2\psi_3+\psi_3\psi_4+
\psi_1\psi_4\cr
&+(1+\ex{-2\beta})(\psi_1\psi_3+\psi_2\psi_4)\cr}} ($\hat\del_x\psi=\psi({\bf
x}+\hat\imath)-\psi({\bf x})$ is the lattice derivative.) The  expansion of
$\int [D\psi]\ex{S}$ reproduces the low-temperature expansion \eZtwob. To see
this, note that we need to bring down four fermions per site in order to get
a nonzero grassmann integral. We can do this in one of two ways: either by
bringing down ``propagator terms", $\psi_1({\bf x})\psi_3({\bf x}+\hat\imath)$
or $\psi_2({\bf x})\psi_4({\bf x}+\hat\jmath)$, each of which comes with a
factor of $\ex{-2\beta}$, or by bringing down ``mass terms" from \eSfermi.
The former must form closed  curves, which get weighted by $\ex{-2\beta
L}$. Sites which are not visited by these curves must have two mass terms
brought
down from the action. There are three different ways of doing this, but by
Fermi
statistics, one gets counted with a minus sign. When the dust settles, we
precisely reproduce the low-temperature expansion \eZtwob.

In the continuum limit \eSfermi\ reduces to the action for a free majorana
fermion of mass $m=(\ex{2(\beta-\beta_c)}-1)/a$ ($a$ is lattice spacing). At
the
critical point, the fermion is massless.
To study the continuum limit, it is helpful to form new linear combination of
the fermions which diagonalize the mass term in \eSfermi\ \DotDot. Let
$\alpha=\ex{i\pi/4}$ and form the linear combinations
 $$\eqalign{
\psi&=\half[\alpha^7\psi_1+\alpha^2\psi_2+\alpha^5\psi_3+\psi_4]\cr
\tilde\psi&=\half[\alpha\psi_1+\alpha^6\psi_2+\alpha^3\psi_3+\psi_4]\cr
\chi&=\half[\alpha^3\psi_1+\alpha^2\psi_2+\alpha\psi_3+\psi_4]\cr
\tilde\chi&=
\half[\alpha^5\psi_1+\alpha^6\psi_2+\alpha^7\psi_3+\psi_4]\quad.\cr
}$$
The action \eSfermi\ becomes
\eqn\eSfermib{\eqalign{S=\sum_{\bf x}\ \ex{-2\beta}\Bigl[&
-{1\over4}(\psi-\chi)\hat\del_x(\psi-\chi)
-{1\over4}(\tilde\psi-\tilde\chi)\hat\del_x(\tilde\psi-\tilde\chi)\cr
&-{i\over 4}(\psi+\chi)\hat\del_y(\psi+\chi)
+{i\over4}(\tilde\psi+\tilde\chi)\hat\del_y(\tilde\psi+\tilde\chi)\cr
&+{i\over4}\bigl(\hat\del_x(\psi-\chi)\hat\del_x(\tilde\psi-\tilde\chi)
+\hat\del_y(\psi+\chi)\hat\del_y(\tilde\psi+\tilde\chi)\bigr)\cr
& +i\hat m_1\psi\tilde\psi-i\hat m_2\chi\tilde\chi\Bigr]\cr}}
where
\eqn\emasses{\hat m_1=(\ex{2(\beta-\beta_c)}-1),\qquad
\hat m_2=2\sqrt{2}\ex{2\beta}+\hat m_1, \qquad \ex{2\beta_c}=1+\sqrt{2}\quad.}
The continuum limit is obtained by letting $\beta\to \beta_c$, while
the lattice spacing $a\to0$, holding $m_1=\hat m_1/a$ fixed. In this limit,
$m_2=\hat m_2/a\sim 2\sqrt{2}/a\to\infty$, so  we can simply drop all
dependence
on $\chi,\tilde\chi$. Looking at the kinetic energy operator that remains for
$\psi,\tilde\psi$, one notices two important features. First, because it is
defined using {\it asymmetric} lattice derivatives, it is non-Hermitian.
In momentum space
$\hat\del_x\psi= \psi({\bf x}+\hat\imath)-\psi({\bf x})\to
2i\ex{ip_x/2}\sin(p_x/2)\psi({\bf p})$, unlike the symmetric lattice derivative
$\half(\psi({\bf
x}+\hat\imath)-\psi({\bf x}-\hat\imath))\to i\sin(p_x)\psi({\bf p})$. Second,
the kinetic energy operator has only one zero in the Brillouin zone, at ${\bf
p}=0$. The latter feature is, of course, what we want. The former is a little
troubling. Generally, it is believed that a non-Hermitian kinetic energy
operator would lead to non-unitarity of the continuum theory. Here it is
clearly harmless. Indeed, expanding about ${\bf p}=0$, one simply replaces
lattice derivatives by continuum derivatives $\hat\del\to a\del$ and, after
rescaling the fields, one obtains
\eqn\escont{S_{cont}=\int \dd^2z\bigl( -\half\psi\del_{\bar z}\psi
-\half\tilde\psi\del_{z}\tilde\psi+im_1\psi\tilde\psi\bigr)}
which is, as promised, the action for a free
majorana fermion of mass $m_1$ ($\del_{z,\bar z}=\half(\del_x\mp i\del_y)$).
All
of the non-Hermiticity of the lattice action went into {\it irrelevant}
operators\footnote{$^\dagger$}{The fact that Wilson-like terms remove the
doubling in the Ising model was noted previously by Hassenfratz and Maggiore
\ref\rhass{P. Hassenfratz and M. Maggiore, Phys.
Lett. {\bf B220} (1989) 215.}.}\
which can be neglected in the continuum limit.

Even global effects are captured correctly. Recall
from our discussion of \potts, that for the fermions to reproduce the Ising
partition function on a lattice with periodic boundary conditions, we need to
introduce independent horizontal and vertical couplings and sum over the
partition functions where we shift the horizontal and vertical couplings by
$i\pi/2$
$$\CZ[\beta]=\half[Z(\beta,\beta)+Z(\beta+i\pi/2,\beta)
+Z(\beta,\beta+i\pi/2)+
Z(\beta+i\pi/2,\beta+i\pi/2)]$$
These shifted partition functions are generated by the modified fermionic
actions
\eqn\eSfermid{\eqalign{S=\sum_{\bf x}\ \ex{-2\beta}(\pm\psi_1\hat\del_x\psi_3
\pm\psi_2\hat\del_y\psi_4)&+\psi_1\psi_2+\psi_2\psi_3+\psi_3\psi_4+
\psi_1\psi_4\cr
&+(1\pm\ex{-2\beta})\psi_1\psi_3+(1\pm\ex{-2\beta})\psi_2\psi_4\cr}}
Going through the same analysis as before, one sees that modifying the action
in
this way changes the boundary conditions for the continuum fermion from
periodic to antiperiodic, so that \eZtwod\ indeed reproduces the sum over spin
structures for the fermion. All of these somewhat subtle features of the
continuum partition function are reproduced by the lattice theory without
doubling or other inconsistencies.

One might think that everything worked out so nicely here because the continuum
limit is a {\it free} theory. This is not so. Consider instead the
Ashkin-Teller
model. This is a theory of two Ising models coupled by a four-spin interaction.
In the fermionic formulation, it consists of two copies of \eSfermi\ (for
$\psi_{1,\dots,4}$ and $\psi_{5,\dots,8}$), coupled by a four-fermi interaction
$$\Delta S=\kappa \bigl((\psi_1\hat\del_x\psi_3 +\psi_1\psi_3)
(\psi_5\hat\del_x\psi_7 +\psi_5\psi_7)+
(\psi_2\hat\del_y\psi_4 +\psi_2\psi_4)(\psi_6\hat\del_y\psi_8 +\psi_6\psi_8)
\bigr)$$
In the continuum limit, this reduces to a massive Thirring
model\footnote{${}^\clubsuit$}{Perhaps, I should say ``Thirring-like" model. It
consists of two massive majorana fermions coupled by a four-fermion
interaction $\lambda \int\dd^2z\ \psi\tilde\psi\psi'\tilde\psi'$.} (massless
along the critical line). This is a genuine interacting field theory.
Interestingly, the {\it location} of the critical point (line) is not shifted
by
the interactions or, to say it differently, the ultra-massive doublers (and the
$\chi$'s) do not feed down their masses to the ``light" fermions through the
interactions. The light fermion masses are protected, not through a continuous
chiral symmetry, but through a discrete symmetry -- the duality symmetry of the
Ashkin-Teller model. This use of a duality symmetry to keep light fermions
light
is a somewhat novel mechanism from the point of view of lattice field theory.
Perhaps it has some wider applicability.

\appendix{B}{Is Nambu-Goto = Polyakov ?}

Most of the comments made in Section 4, tacitly assumed that one knew, at least
in principle, how to quantize the string theory. For the Nambu-Goto string, we
really know very little, but it is usually assumed that Nambu-Goto is
equivalent to the Polyakov string.

The usual argument \Polybook\ proceeds as follows. Introduce an intrinsic
metric $g=f^*(\ex{\phi}\hat g)$ on the surface $\Sigma$, where $\hat g$ is a
fiducial metric and $f$ a diffeomorphism. Rewrite the Nambu-Goto action as
\eqn\nambu{\int \sqrt{g} +i \lambda^{ab}(\del_a{\bf
X}\cdot\del_b{\bf X} -g_{ab})}
where $\lambda^{ab}$ is a lagrange-multiplier field which sets the intrinsic
metric $g_{ab}$ equal to the induced metric
$h_{ab}=\del_a{\bf X}\cdot\del_b{\bf
X}$. (I have written the intrinsic metric in the conformal gauge
because that's where I understand best how to quantize the theory, but the
following remarks hold true in the light-cone gauge as well.)

If we assume that the lagrange-multiplier field $\lambda$ has short-range
correlations, then one can solve for it using the equations of motion, in which
case, \nambu\  becomes equal to the Polyakov action.

Unfortunately, for a 3-dimensional target space, the generic singularity of the
induced metric $h$ has {\it one} of its eigenvalues going to zero, while the
other remains finite. We would not be too upset if both of the eigenvalues went
to zero (as happens for a 1-dimensional target space); that would simply
correspond to $\phi\to -\infty$, which is simply the boundary of field-space
and
is ``included" in the $\phi$ functional integral. But this Whitney-type
singularity \whitney, where one eigenvalue goes to zero, does not correspond to
{\it any} configuration of the field $\phi$. It is simply inaccessible in
conformal gauge. (More properly, it is related by a {\it singular}
``diffeomorphism" to a metric in the conformal gauge
slice.)\footnote{${}^\star$}{ Sedrakyan\sedratalk\ has
argued that one might try to mimic the effect of these Whitney singularities
by a gas of operator insertions on an otherwise smooth surface. In the most
na\"\i ve approach, operator insertions simply produce ordinary conical
singularities, which is not what we want.}

This should make one a little worried (but perhaps only a little) about the
equivalence of between the Nambu and Polyakov actions for D=2,3, where the
generic singularity is of the Whitney type.

\appendix{C}{Fermions on nonorientable manifolds}

Because of the difficulties alluded to in the text, it must be the case that
the definition of fermions used in
open string theory {\it must} be different from that studied by Grinstein and
Rohm \rohm. In particular, the m\oe bius band, viewed in the closed string
channel, has both a NS-NS sector and an R-R sector, whereas the fermions
discussed in \rohm\ have only an R-R sector.
Here I will explain the discrepancy: there are two {\it inequivalent}
definitions of fermions on a nonorientable manifold.

 The pinor bundle has a
structure group which is a $\BZ_2$ extension of the structure group of the
manifold
\eqn\Oext{0\to\BZ_2\to Pin(n)\to O(n)\to 0}
The m\oe bius band and the Klein
bottle (the only nonorientable manifolds for which explicit calculations have
been done by open string theorists) are both flat 2-manifolds. Thus the
structure group can be reduced from $O(2)$ to $\BZ_2$. There are
clearly two possible $\BZ_2$ extensions of the reduced structure group: $\BZ_4$
or $\BZ_2\times\BZ_2$. The definition of fermions used by open string theorists
corresponds to the former (the transition functions of the fermions are in
$\BZ_4$), whereas the fermions in \rohm\ correspond to the latter extension.

It is not immediately evident whether the former definition can be extended to
curved 2-manifolds with the full $O(2)$ structure group. As it turns out, it
can \CBDM
\footnote{${}^\star$}{I'd like to than R. Rohm for bringing this
reference to my attention.}. Here I will just summarize the results, refering
to \CBDM\ for details.

One starts with the Clifford algebra $\CC^\pm_n$ (over $\BR$), generated by the
Dirac matrices $\gamma_a$ with the relation
\def\cliff{(2)}
\eqn\cliff{\gamma_a\gamma_b+\gamma_b\gamma_a=\pm2\delta_{ab}\BI}
On $\CC^\pm_n$, one defined a ``norm" by
$$\|\gamma_{v_1}\dots\gamma_{v_N}\|^2=|{\rm Tr} u^2|$$
(where ${\rm Tr}$ is the usual quadratic form on the Clifford algebra,
normalized so that ${\rm Tr} \BI=1$) and the group $Pin^\pm(n)$ is the
multiplicative group of invertible elements of $\CC^\pm_n$ of unit norm. In
each
case, $Pin(n)$ is the extension by $\BZ_2$ of $O(n)$, but in general
$Pin^+(n)\not\simeq Pin^-(n)$. In particular, consider $Pin(1)$, which consists
of 4 elements, $\{\pm\BI,\pm\gamma\}$.
But depending on whether $\gamma^2=\pm\BI$,
we obtain different groups: $Pin^+(1)=\BZ_2\times\BZ_2$,
while
$Pin^-(1)=\BZ_4$.

$Spin^\pm(n)$ is
the subgroup of $Pin^\pm(n)$ which extends the identity component of $O(n)$
$$0\to \BZ_2\to Spin^\pm(n)\to SO(n)\to 0$$ However, one always finds that
$Spin^+(n)\simeq Spin^-(n)$. Thus {\it for orientable manifolds} the choice of
sign in \cliff\ turns out to be irrelevant to defining spinors. However, for
nonorientable manifolds, {\it it matters}!

We then go on to try to use the group extension \Oext\ to lift the transition
functions of the $O(n)$ frame bundle on $M$ to those of a principal
$Pin^\pm(n)$ bundle. In so doing, we encounter an obstruction. Even though
the $O(n)$ transition functions obey the cocycle condition
$$g_{ij}\cdot g_{jk}\cdot g_{ki}=\BI\quad,$$
when lifted to $Pin(n)$ transition functions, they may not:
$$\tilde g_{ij}\cdot \tilde
g_{jk}\cdot \tilde g_{ki}\in\BZ_2\quad .$$
This defines an element of $\H{2}{M,\BZ_2}$.
For $Pin^+(n)$, the obstruction is the second Stiefel-Whitney class $w_2(M)$.
For $Pin^-(n)$, the obstruction is $w_2+w_1\cup w_1$. The former, as noted in
\rohm, is nonvanishing on certain 2-manifolds. Hence one cannot define a
$Pin^+$ bundle on those manifolds. The latter (as one sees from equations
\einter,\eSW) {\it vanishes identically} on any 2-manifold. Thus, we can
always define a $Pin^-$ structure on any 2-manifold and this (not the more
intuitive $Pin^+$) is the definition of fermions used by open string
theorists\footnote{${}^\dagger$}{The $SO(32)$ open string defined using
$Pin^+$  is a rather amusing variant. It is an anomaly-free string theory with
a nonvanishing cosmological constant (and hence a dilaton tadpole). In that
respect, it resembles the $SO(16)\times SO(16)$ heterotic string. The theory is
anomaly free because the m\oe bius band and cylinder amplitudes in the R-R
sector are exactly the same as in the usual $SO(32)$ open string. However the
dilaton tadpole does not cancel between the disk and $\BR P^2$, as it does in
the usual $SO(32)$ string. Rather, the dilaton 1-point function on $\BR P^2$ --
defined physically by sewing -- vanishes identically.}.

\bigbreak\bigskip\bigskip\centerline{{\bf Acknowledgments}}\nobreak
\frenchspacing{
I would like to thank M. Bershadsky, P. Ginsparg, D. Kutasov, A. Polyakov,
and H. Verlinde for useful conversations. I would especially like to thank A.
Sedrakyan, conversations with whom kindled my interest in this subject.
This work was supported by
NSF grant PHY90-21984. }

\listrefs
\end